\documentclass[runningheads]{llncs}
\usepackage{url}

\usepackage{xcolor}
\usepackage{array}

\newcommand{\private}[0]{\textit{PrivateCompany}}

\newcommand{\inhouse}[0]{\textit{InhouseCompany}}

\begin{document}
\mainmatter              
\title{Does Ownership Structure Matter?\\
A Case Study on Business Performance\\ of Two Accounting Companies} 
%
\titlerunning{Does Ownership Structure Matter?}  
%
\author{Reetta Ghezzi\inst{1} \and Sanni Marjamäki \inst{1} \and Teemu Laine \inst{2} \and Tatu Virta \inst{} \and Hannu Vilpponen \inst{1} \and Tommi Mikkonen \inst{1}}

\authorrunning{Ghezzi et al.} 
%
\tocauthor{Reetta Ghezzi, Sanni Marjanen, Teemu Laine, Tatu Virta, Hannu Vilpponen, and Tommi Mikkonen}
\institute{University of Jyväskylä, Finland,\\
\email{reetta.k.ghezzi@jyu.fi}
\and
Tampere University, Finland}

\maketitle              

\begin{abstract}
Many public organisations procure a substantial amount of goods and services from in-house companies. When providing their goods and services, those companies are supposed to fulfil objectives set for them and for the wider entity, including in particular cost-effectiveness. This paper examines the performance of selected in-house companies both by analyzing the externally reported financial performance (top-down) and analyzing the internal operations and related performance (bottom-up). Based on the analysis, it is discussed, 1) how the in-house companies fulfil their assigned tasks as publicly owned entities and 2) how these in-house companies and their performance should be controlled by public bodies. Methodologically the paper takes advantage of two cases of accounting companies: one publicly owned in-house company and another private company. As a conclusion, in this case top-down analysis reveals inefficiencies which are further explored via bottom-up analysis. In this case, privatization leads to enhanced business performance.

\end{abstract}
%

\section{Introduction}


In Finland public sector bought goods and services worth EUR 45 billion in 2023 \cite{tutki:hallintoa}. Public procurement plays a vital role in the economic activities of government entities, ensuring that the acquisition of goods and services is conducted transparently, efficiently, and fairly. The Public Procurement Act governs public acquisitions, but it includes specific provisions for in-house procurement. Simply put, in-house procurement allows contracting authorities, such as municipalities, to procure goods and services directly from companies they own without tendering \cite{hankinta:laki}.
 
It has been evaluated, that Finnish public agencies own approximately 2700 in-house companies \cite{inhouse:maara}. The in-house market is approximately EUR 5-10 billion euros \cite{pyykkonen:saastokeinot}. The latest review on in-house procurement in Finland approximates the in-house market size to be EUR 3,7 billion \cite{jaaskelainen:kkv}, however, this figure does not include in-house companies in sectors such as energy, water and real estate. Therefore, the previous approximation is more realistic. 

In-house procurement aims at streamlining processes and at reducing costs for public entities. However, questions have been raised about its effectiveness, the potential for market distortion, and whether contracting authorities genuinely exert control over their in-house companies \cite{KKV:marketcourt}. The efficacy of in-house companies is further scrutinized in various European contexts, where countries have implemented stringent criteria to ensure alignment with market efficiency and public interest. In Finland, concerns about market distortion and the effective use of public funds persist. The Confederation of Finnish Industries and the Finnish Competition and Consumer Authority have raised these issues, advocating for stricter oversight and transparency in in-house procurement practices. Programme of Prime Minister Petteri Orpo's Government \cite{government:programme} has raised the in-house procurement legislation under re-evaluation by presenting a minimum of 10\% ownership over one's in-house company. This means approximately fifty in-house companies in Finland, mainly in the sectors of financial and payroll management and ICT would encounter rearrangements \cite{jaaskelainen:kkv}.

This research aims to contribute to practice by exploring in-house and traditional companies through business performance metrics. By analyzing key financial indicators such as hourly rates, labour costs, and operational efficiency, the study exposes inefficiencies tied to resource utilization and cost management. The change in ownership structure, from in-house to private (privatization), serves as the observed phenomenon that reveals these inefficiencies. This shift allows for a deeper understanding of how business performance is affected, providing valuable insights for public sector organizations seeking to enhance efficiency and optimize resource management. The research questions guiding this study are:
\begin{itemize}
    \item RQ1: How effective are in-house companies in fulfilling their intended roles of reducing costs and improving service delivery for public entities?"
    \item RQ2: How does privatization affect business performance?
\end{itemize} 

In this context, the research will explore the intricacies of in-house procurement, providing a comprehensive understanding of its potential benefits and drawbacks. The aim is to contribute to the ongoing discourse on public procurement by offering insights into the effectiveness and governance of in-house companies, thereby aiding policymakers and public entities in making informed decisions about their procurement strategies.

The rest of the paper is structured as follows. In Section 2, we provide background and motivation for the work. In Section 3, we describe the research method applied in this paper. In Section 4, we present the results. In Section 5, we discuss the implications of this work. Finally, in Section 6, we draw some conclusions.
\section{Background and Motivation}

\subsection{Defining In-House Entity}


The Public Procurement Act \cite{hankinta:laki} governs public acquisitions. However, it does not apply when a contracting authority, for example, a municipality, makes a procurement from a company it owns, called an in-house entity or in-house company, provided that the in-house company is formally separate for policy-making purposes, has a controlling interest by the municipality and conducts only a limited amount of business with external parties \cite{hankinta:laki}. Procurement Directive allows 20 per cent of turnover to go outside the owners of the in-house company \cite{hankinta:direktiivi}. However, in Finnish law, the threshold for outselling is stricter. Public Procurement Act specifies that 5 per cent and EUR 500,000 limits for outselling apply based on the in-house entity's turnover three years before the agreement \cite{hankinta:laki}. However, these limits don't apply when there's no market-based operation to execute the services. Whether the market-based operations exist is determined by the responses to a transparency declaration \cite{hankinta:laki}.

Procurement units that own the in-house company must have decisive authority in the in-house company \cite{hankinta:direktiivi}. The Public Procurement Act defines joint-decisive authority as when all contracting entities have representatives in the in-house company's executive organs and collectively make strategic decisions, with the condition that the in-house company operates in the interests of the controlling contracting entities \cite{hankinta:laki}. In addition, the Public Procurement Act states that it does not apply when an in-house company is a procurement unit itself and procures goods or services from another procurement unit, which exercises controlling interest in the in-house company or another entity under the same controlling interest \cite{hankinta:laki}. This option is the so-called in-house sisters' arrangement in Finland. 

\subsection{Diving into the In-House Dynamics}

Recent public discussion has been raised over the in-house position as habitual practice through ownership and a fictitious demonstration of decisive authority. The question of decisive authority in-house entities was examined in the market court decision MAO:154/2024 \cite{MAO:vantaakerava}. In this case, a Wellbeing Services County (a regional public body for healthcare and social services in Finland) procured human resource management services and related systems directly from Sarastia without competitive bidding, under an in-house procurement valued at approximately €9.3 million. The welfare area owned only 0.04 per cent of Sarastia's shares.

The market court concluded that the welfare area did not exert actual control over Sarastia, either alone or together with other shareholders. The court found that the welfare area's ability to influence Sarastia's operations through share ownership or any of Sarastia's governing bodies was minimal. Consequently, Sarastia could not be considered an in-house entity of the Wellbeing Services County, and the procurement made by the Wellbeing Services County from Sarastia should not be exempt from the public procurement law \cite{KKV:marketcourt}.

Based on these legal cases, it is evident that the importance of real decisive authority and ownership in the in-house company is significant. In addition to ownership share, importance is also given to control, structure, decision-making, and genuine representation in the in-house company's operations. It is important to assess these factors as a whole when evaluating the legal status of an in-house company. To overcome the issues mentioned above, for example, Halonen \cite{halonen:inhouse} describes that Italy has ruled that one should not bind in-house relationships if the decisive authority over the upcoming in-house entity is not guaranteed. Currently, in Finland, when carefully estimated, we are allowing fifty such in-house companies to operate which may or may not have adequate in-house status. 

In practical applications, in-house procurement may benefit smaller municipalities by reducing the bureaucracy involved in contracting and contract implementation costs \cite{miemec:2019}. However, it has been questioned whether the upcoming, now-current directives will create a procurement market that does not have to obey and is not controlled by procurement norms \cite{hausmann2013house}. The concerns eleven years ago were that the procurement directives would exclude private service providers from the competition if the in-house exception is accepted as it was drafted \cite{hausmann2013house}. Today, we can depict, that these concerns were accurate. 

Similar concerns are still present in Finland and maybe in other EU countries as well. The Confederation of Finnish Industries has raised concerns that the current in-house practice distorts the market and has taken steps to address these concerns through a request for measures to the practices from the Competition and Consumer Authority \cite{EK:2024}. Baciu \cite{baciu2015horizontal} suggests that public bodies should not be able to avoid transparent procedures and contract directly with other public bodies, except in rare and limited situations to preserve fair competition \cite{baciu2015horizontal}.  The Confederation of Finnish Industries and the Finnish Competition and Consumer Authority also take the same view in their proposals \cite{EK:2024,KKV:esitys}. Yet again, Halonen \cite{halonen} depicts, that in Lithuania, in-house procurement is only allowed if the market cannot offer goods and services, which has been debated in Finland as well. 

The literature concludes the current procurement directive inhibits opening up the national procurement markets and fosters direct awarding in public contracts, even if the underlying purpose is the opposite. In our research, the emphasis is to overall evaluate the efficiency of the in-house companies. 


\subsection{Exploring the performance measurement theme}

The business performance analyses can be conducted at the firm level, based on the externally reported financial statements, and based on that overall financial information, more detailed observations can be made, for instance regarding the profitability, liquidity and solvency of the firm. The observations can be made with the help of performance indicators describing those perspectives. Also, regarding them, comparisons to targets, industry averages and other companies can be made. Also, trends regarding them can be helpful in analyses. The approach, where company-level analyses lead into more detailed observations, here regarding inhouse company business performance, is called here as “top-down” approach.
 
Another alternative to understanding business performance is to depart from the business activities, processes and individual goods and services. All the business activities are in place to make sure that the company is able to provide its customer(s) with goods and services, also in the long run. In any multi-product company, the profitability between different product varies greatly, as the pricing is not directly based on costs, and as the companies do not necessarily know their current costs or their dynamics. Activity-based costing \cite{Cooper:1992,Laine:2012} helps allocating cost of resource use to activities, processes and eventually to cost objects such as goods and services provided to the customers. In the context of in-house companies, it is important to know how the activities and processes match the requirements of the goods and services and eventually fulfil the aims to serve the customers efficiently and effectively. The approach, where the starting point is the internal processes and activities and their performance is here called bottom-up approach for the in-house company business performance.
 
Also, by combining these two approaches is useful, as the overall performance may raise questions that more detailed analyses on the activities could answer. On the other hand, by combining the analyses on individual activities and processes, one should get a picture that aligns with the overall performance. If this is not the case, such a comparison may reveal perspectives for further analysis.

\subsection{The Case}

 \private \ specializes in comprehensive financial management services, including bookkeeping, payroll administration, financial automation, tax consultation, business restructuring, and collection services. They offer tailored solutions for businesses and public agencies of all sizes, with a focus on personalized service, efficiency, and sustainability.


\inhouse's parent company has 285 shareholders, which include municipalities, municipal companies, and wellbeing services counties. The ownership structure ensures that only the shareholders of the parent in-house company can access the services provided by daughter in-house companies. This exclusive service provision is a strategic approach to ensure that the resources and benefits generated within the group are retained among its stakeholders.

\inhouse \ extends its service offerings beyond the group's shareholder base. Unlike the parent in-house company, which is restricted to serving only its shareholders, \inhouse \ provides its financial and human resource management services to clients who are not shareholders. This unique position allows \inhouse \ to operate in a broader market, potentially increasing its client base and revenue streams while still being the in-house entity for the parent in-house company.

\inhouse's primary services in 2022 included financial and human resources management, with substitute services forming the bulk of the company's volume from both financial and HR perspectives. The year 2022 marked \inhouse \ fourth year of operation. Its accounting and payroll services, which were transferred to \private \ on November 30, 2022, through a business transaction, accounted for approximately €380,000 of its revenue. This transition reflects the strategic acquisition by \private \ to enhance its service portfolio.



\section{Research Method}

In researching the change in ownership structure from in-house company to private company, the case study method provides possibility to use dual framework through which the transition can be analyzed, using both top-down and bottom-up approaches. 

The top-down approach initiated by examining the company's overall financial performance, leveraging external financial statements to understand profitability, liquidity and solvency. This method enables identifying broad trends and compare the firm's financial outcomes to industry benchmarks and targets. Applied to the ownership transition, this approach would focus on how financial health of the company evolves as it moves from public to private ownership. For example, does privatization improve profitability due to streamlined operations or reduced public sector restrictions? Or do liquidity and solvency metrics shift due to changes in capital structure and funding sources? This perspective offers insights into the strategic and financial implications of the ownership shift, helping to assess whether the transition aligns with the broader corporate goals.

On the other hand, the bottom-up approach focuses on internal activities, processes, and the performance of individual products or services. In this context, the method delves into how operational processes are impacted by the transition. The focus shifts to how efficiently and effectively resources are used post-privatization. Activity-based costing, for example, can be employed to allocate costs to specific processes and services, shedding light on operational changes under private ownership. This method helps to assess whether privatization enhances the alignment between internal processes and the company’s ability to meet customer demands. Are there improvements in cost management or service quality after the change? Or do inefficiencies emerge as the company adapts to a private-sector business model? The bottom-up approach emphasizes the importance of understanding the operational impact of ownership change on the company's internal mechanics. 

Integrating these two approaches can achieve a more comprehensive understanding of the ownership change. The top-down analysis might reveal overall performance improvements or concerns that require deeper investigation. The bottom-up analysis, focusing on activities and processes, helps explain why certain financial outcomes occur. For example, if profitability increases but certain internal processes show inefficiency, it may suggest areas for further optimization post-privatization. Conversely, strong internal alignment but weak financial performance may highlight strategic misalignments that need attention. These perspectives create a holistic picture of the transition from in-house to private ownership, ensuring that strategic and operational dimensions are fully explored.
This chapter outlines the research methodology employed in this study, focusing on the specific features of the case study approach. The primary methods of data collection included document analysis, interviews, and fact-checking through emails. These methods were selected to ensure a comprehensive understanding of the differences in employee costs, particularly through the analysis of service agreements and collective bargaining agreements. In addition, top-down analysis from financial statements was utilized to understand the financial state of these organizations. 

 The case study approach was chosen for its ability to provide an in-depth understanding of complex issues within their real-life context \cite{myers:2008}. This method is particularly suitable for exploring the detailed and multifaceted nature of service business development and the associated management accounting practices. The case study focuses on specific organizations, \private \ and \inhouse, examining their practices and the impact of different collective bargaining agreements on employee costs.

Document analysis was a critical component of this research. Various documents were collected and analyzed to understand the intricacies of service agreements and their financial implications. The documents reviewed included:

\begin{itemize}
    \item Financial Agreements and Action Reports to understand the financial status these organizations hold.
    \item Service Agreements: Detailed contracts between \inhouse \ and its clients were analyzed to identify terms related to labor costs, service delivery, and financial arrangements. Contracts from \private \ with similar services were then compared to \inhouse \ agreements.
    \item Collective Bargaining Agreements: Both the ERTO agreement used by Accounting Office \private \ and the AVAINTA agreement used by \inhouse \ were scrutinized to pinpoint differences in wage structures, benefits, and other employee-related costs.
    \item Internal Reports and Memos: These documents provided insights into the company’s internal cost management practices and strategic decisions as well as problems that had arisen before business acquisition.
\end{itemize}

An interview with CEO was conducted to gather qualitative data and gain deeper insights into the organizational context and management perspectives. The CEO of \private \ was interviewed to understand the strategic rationale behind their use of specific collective bargaining agreements and their impact on employee costs. Key aspects of the interview process included. Emailing was employed as a method to check facts and clarify information gathered from other sources. This method ensured the accuracy and reliability of the data. Follow-up emails were used to gather additional information or clarification as needed.

A thorough analysis of the collective bargaining agreements was conducted to compare the ERTO and AVAINTA agreements. This analysis focused on:

\begin{itemize}
    \item Wage Structures: Comparison of base salaries, hourly wages, and overtime rates.
    \item Benefits: Examination of holiday compensation, sick leave policies, and other employee benefits.
    \item Cost Implications: Assessment of the overall financial impact of each agreement on the company’s labour costs.
\end{itemize}
    
The data collected through document analysis, interviews, and email correspondence were analyzed using qualitative methods. Key steps in the data analysis process included comparative analysis meaning that data from different sources was compared to identify patterns and discrepancies. Triangulation, meaning that the information from multiple sources was cross-verified to ensure the validity and reliability of the findings.

\begin{table}[]
\begin{tabular}{lllllll}
\hline
                    & Client A      & Client B & Client C & Client D & Client E & Client F \\ \hline
Payroll Services                      & x             & x        & x         &  x       &          &         \\
Efficiency in Customer \\ Services  & x        &      & 
  &                 & x        & x        \\
Collective Bargaining \\ Agreements      & x             & x        & x        & x        &         &         \\
Service Agreements                    & x             & x        & x        &  x        &          &        \\ \hline
\end{tabular}
\end{table}

In terms of confidentiality, the sensitive information was anonymized to protect the privacy of the clients and the organizations. Findings were shared with the participants to ensure accuracy and provide them with an opportunity to offer feedback.
\section{Results}


\subsection{Top-down analysis with Financial Comparisons and Profitability Analysis}

\private \ and \inhouse \ present contrasting financial profiles based on the provided data for 2022. \private \ generated a revenue of €2,366,860.60 with 25 employees, resulting in a revenue per employee of €94,674.42. The company maintained a labour cost-to-revenue ratio of 46.53\%, indicating efficient management of personnel expenses relative to its income. The operating profit for \private \ was €313,011.38, contributing to a solid net profit of €250,413.29 for the fiscal year. These figures suggest a stable and profitable operation with a strong ability to manage costs and generate returns.

\begin{table}[t]
\caption{Some key figures of case organizations.}
\centering
\begin{tabular}{lll}
\hline
\multicolumn{1}{c}{\textbf{Indicator}}      & \multicolumn{1}{c}{\textbf{\private}} & \multicolumn{1}{c}{\textbf{\inhouse}} \\ \hline
Revenue (€)                                 & 2,366,860.60                                             & 11,403,309.00                          \\
Number of Employees                         & 25                                                       & 15                                     \\
Revenue per Employee (€)                    & 94,674.42                                                & 760,220.60                             \\
Labor Costs as a Percentage of Revenue (\%) & 46.53\%                                                  & 90.99\%                                \\
Operating Profit (€)                        & 313,011.38                                               & 419,756.42                             \\
Net Profit/ Loss for the Fiscal Year (€)     & 250,413.29                                               & 336,743.18                             \\ \hline
\end{tabular}
\end{table}

In contrast, \inhouse \ reported a significantly higher revenue of \\ €11,403,309.00 with only 15 employees, leading to an exceptionally high revenue per employee of €760,220.60. However, \inhouse \ employed 2036 temporary workers in 2022 which is why the revenue per employee is not a suitable number for comparison in this case. Therefore, the labour costs are a better indicator. Labour costs accounted for 90.99\% of the revenue. This highlights a substantial expense to income, which impacts overall profitability. 

\inhouse \ demonstrates strong revenue generation, with significant gross margin and contribution margin figures. However, its high labour cost percentage suggests that cost management is less efficient compared to \private , which has tighter control over its expenses, making it more cost-effective despite smaller absolute profits.

\inhouse's operating profit was €419,756.42, which is a substantial amount. However, the efficiency of converting revenue into operating profit is reduced by higher labour costs and reliance on public funds. In contrast, \private \ demonstrates better cost management, turning a smaller revenue base into a more stable operating profit margin.

The net profit for \inhouse \ was €336,743.18, a higher absolute figure than that of \private. However, \private's better profitability metrics relative to its size highlight more efficient financial performance. The disparity in net profit between the two companies stems from differences in how efficiently they manage operational costs, especially labor.

While \inhouse \ excels in generating absolute profits, its ROE might be impacted by its reliance on public funding and the less efficient cost structure. On the other hand, \private \ has a higher ROE due to its more efficient use of equity to generate profit.

\private's acquisition of \inhouse's accounting and payroll services further solidified its market position, giving it a higher equity ratio, which contributes to its financial stability. \inhouse, with its higher debt levels, depends more on external funding sources, which could pose risks to financial sustainability if public funding policies change.

Finally, \private's quick ratio reflects a stronger liquidity position, further enhancing its ability to manage short-term financial obligations efficiently. \inhouse's quick ratio is comparatively weaker, adding another dimension of financial risk tied to its reliance on public sector clients and public funding.

\begin{table}[]
\begin{tabular}{lll}
\hline
\textbf{Indicator}                  & \multicolumn{1}{c}{\textbf{\private}} & \multicolumn{1}{c}{\textbf{\inhouse}} \\ \hline
\textbf{Gross Margin (€)}           & €1,820,446.64                                            & €10,882,196.87                          \\
\textbf{Gross Margin (\%)}          & 76.91\%                                                  & 95.43\%                                \\
\textbf{Contribution Margin (€)}    & €326,364.05                                              & €422,190.98                             \\
\textbf{Contribution Margin (\%)}   & 13.79\%                                                  & 3.70\%                                 \\
\textbf{Operating Profit (\%)}      & 13.22\%                                                  & 3.68\%                                 \\
\textbf{Net Profit (\%)}            & 5.54\%                                                   & 1.27\%                                 \\
\textbf{ROE (Return on Equity, \%)} & 8.05\%                                                   & 22.41\%                                \\
\textbf{Equity Ratio (\%)}          & 72.00\%                                                  & 37.90\%                                \\
\textbf{Relative Indebtedness (\%)} & 27.11\%                                                  & 9.28\%                                 \\
\textbf{Quick Ratio}                & 4.18                                                     & 1.60                                   \\ \hline
\end{tabular}
\end{table}

\private \ is more reliable and efficient when examining the profitability indicators. With an equity ratio of 72.00\%, \private \ demonstrates a robust financial structure, suggesting a strong ability to finance its operations and growth with internal resources. Its relative indebtedness of 27.11\% indicates a manageable level of debt, further supporting its financial stability. In terms of profitability, \private \ shows a contribution margin of 13.79\%, a gross margin of 76.91\%, and a net profit margin of 5.54\%, all of which point to efficient cost management and a solid ability to convert revenue into profit. Additionally, its return on equity (ROE) of 8.05\% signifies a decent return on shareholders' investments, while the quick ratio of 4.18 highlights excellent liquidity, indicating the company’s strong capacity to cover short-term liabilities.

On the other hand, \inhouse \ demonstrates a lower equity ratio of 37.90\% and a lower relative indebtedness of 9.28\%, suggesting less reliance on debt but also a weaker financial backbone compared to \private. Despite having a high gross margin of 95.43\%, \inhouse’s contribution margin stands at only 3.70\%, and its net profit margin is a mere 1.27\%, reflecting lower overall profitability and efficiency. However, \inhouse's return on equity (ROE) is notably high at 22.41\%, indicating strong profitability from the perspective of equity investors. Yet, the quick ratio of 1.60, while adequate, is significantly lower than that of \private, suggesting more limited liquidity.

In summary, while \inhouse \ excels in terms of sales efficiency and ROE, \private's stronger financial structure, higher profitability margins, and superior liquidity position make it the more reliable and efficient company based on the provided financial indicators. However, concerns arise regarding \inhouse's reliance on public funds and its efficiency in cost management. The high labour cost percentage indicates potential inefficiencies, and the dependency on public funding may pose risks to financial sustainability, especially if there are changes in public funding policies or allocations. This reliance underscores the need for vigilant financial oversight and strategic planning to ensure long-term stability and compliance with public sector funding requirements.

\subsection{Automation and Changes in Work Practices}
\private \ supports both traditional and electronic financial management methods, offering real-time financial tracking and electronic invoicing. Their advanced payroll management software, including Mepco, Netvisor, and Fennoa, streamlines payroll processes, automates calculations, ensures timely payments, and maintains comprehensive employee data. These tools enhance efficiency, reduce errors, and free up time for core business activities, ensuring secure, reliable, and efficient financial management. 

The transition from \inhouse \ to \private \ has led to significant improvements in work practices and processes. This change has not only streamlined operations but has also resulted in measurable financial and operational benefits. One of the most notable outcomes is the enhancement of work practices and processes, which has contributed to an overall billing increase of 22\%. This indicates that the organization has become more efficient in managing its financial operations and client billing cycles.

Furthermore, the efficiency per employee has seen a remarkable improvement of 82.5\%. This substantial increase in individual productivity highlights the effectiveness of \private's processes and the successful implementation of more efficient work methods. The reduction in hours required for clients is a clear indicator of this enhanced efficiency. Furthermore, after the business aqcuisition \private \ transitioned the pricing model from hourly rate basis to a monthly subscription model. This change incentivizes more efficient processes. For example, for Client D, the hours used by \inhouse \ were 150 hours, whereas \private \ used only 124 hours. Similarly, for Client E, \inhouse's hours amounted to 240, while \private's hours were significantly reduced to 105. In addition, the transition from \inhouse \ to \private \ has significantly leveraged automation and updated skills to enhance operational efficiency. This modernization has led to a 55\% improvement in client service efficiency. By integrating advanced automated systems and ensuring continuous skill development, \private \ has optimized its processes, reduced manual effort, and increased the accuracy and speed of service delivery.

\subsection{Bottom-up analysis}

\begin{table}[t]
\caption{Employee related costs in case organizations.}
\centering
\footnotesize
\begin{tabular}{llll}
\hline
\multicolumn{1}{c}{\textbf{Indicator}} & \multicolumn{1}{c}{\textbf{\private}} & \multicolumn{1}{c}{\textbf{\inhouse}} & \multicolumn{1}{c}{\textbf{Difference}} \\ \hline
Monthly Salary                         & €2,077.0                            & €2,231.0                            & 7\%                                        \\
Holiday Compensation \\ to Employee       & €3,282.0                            & €3,837.0                            & 14\%                                       \\
Holiday Replacement Cost \\ to Company    & €2,596.0                            & €3,824.0                            & 32\%                                       \\
Annual Salary Cost \\ to Company          & €28,405.0                           & €30,726.0                           & 8\%                                        \\
Annual Working Hours                   & 1,762.5                             & 1,624                               & -9\%                                       \\
Difference in Hourly Labor Cost        & €16.1                               & €18.9                               & 15\%                                       \\ \hline
\end{tabular}

\end{table}

\subsubsection{Analysis of Wage and Labor Cost Comparison}
The table provides a detailed comparison of wages and labor costs between \private \ and \inhouse, based on the ERTO collective bargaining agreement used by \private \ and the AVAINTA collective bargaining agreement used by \inhouse. The data reveals several significant differences that have implications for the efficiency and financial management of these companies, particularly given \inhouse's reliance on public funds.

Employees at \inhouse \ receive a higher monthly salary (€2,231.0) compared to those at \private \ (€2,077.0), indicating a 7\% difference. This suggests that the AVAINTA collective bargaining agreement provides more favorable base salary terms for employees. Additionally, holiday compensation is higher at \inhouse \ (€3,837.0) than at \private \ (€3,282.0), reflecting a 14\% difference. This points to potentially more generous holiday benefits under the AVAINTA agreement, which might improve employee satisfaction but also increases costs.

The cost of hiring replacements during holidays is significantly higher for \inhouse \ (€3,824.0) compared to \private \ (€2,596.0), with a substantial 32\% difference. This indicates that operational costs during holiday periods are more burdensome for \inhouse, possibly due to higher wages or additional benefits required by the AVAINTA agreement. Furthermore, the annual salary cost to the company is higher for \inhouse \ (€30,726.0) compared to \private \ (€28,405.0), reflecting an 8\% increase. This suggests that the overall cost of maintaining an employee is greater for \inhouse, which weakens its financial efficiency.

In terms of working hours, employees at \private \ work more hours annually (1,762.5 hours) than those at \inhouse \ (1,624 hours), indicating a 9\% reduction in hours at \inhouse. This could imply that the AVAINTA agreement provides for shorter working hours or more leave entitlements. Additionally, the hourly labor cost is higher at \inhouse \ (€18.9) compared to \private \ (€16.1), indicating a 15\% difference. This highlights higher per-hour employment costs at \inhouse, which affects the company’s labor cost structure and overall profitability.

Given that \inhouse \ is publicly funded, these cost disparities raise concerns. The expectation when separating functions from public agencies to firms owned by public entities is to enhance efficiency and reduce costs, not increase them. However, the data indicates that \inhouse’s operational costs, particularly in terms of labor, are higher than those of \private. This could undermine the goal of cost-efficiency and calls for a careful evaluation of how public funds are being utilized.

For \inhouse, which relies on public funding, the higher costs associated with the AVAINTA collective bargaining agreement necessitate vigilant financial oversight and strategic planning. The aim should be to ensure that the transfer of functions from public agencies to publicly owned firms results in improved operational efficiency and cost savings, rather than increased financial burdens. Effective management practices and potential renegotiation of labor agreements could be considered to align \inhouse’s operational costs with the efficiency goals expected of publicly funded entities.

\begin{table}[t]
\caption{Services and their costs in case organizations.}
\centering
\begin{tabular}{llll}
\hline
\multicolumn{1}{c}{\textbf{Service}}  & \multicolumn{1}{c}{\textbf{\private}} & \multicolumn{1}{c}{\textbf{\inhouse}} & \multicolumn{1}{c}{\textbf{Difference \%}} \\ \hline
Payroll processing, \\ cost per employee & €26.43                              & €85.00                              & 222\%                                      \\
Bookkeeping                           & €64.90                              & €85.00                              & 31\%                                       \\
Demanding expert work                 & €89.00                              & €125.00                             & 40\%                                       \\
Work outside the agreement            & €64.90                              & €135.00                             & 108\%                                      \\
Assistant work                        & €60.00                              & €89.00                              & 48\%                                       \\ \hline
\end{tabular}
\end{table}

\subsubsection{Comparing Hourly Rates}
In the 2020 agreements, there is a notable disparity in hourly rates between \private \ and \inhouse. For payroll processing, the cost per employee at \private \ is €26.43, while at \inhouse \ it is significantly higher at €85.00, representing a 222\% difference. Bookkeeping services also show a discrepancy, with \private \ charging €64.90 per hour compared to \inhouse's €85.00, a 31\% difference. For demanding expert work, \private's rate is €89.00 per hour, whereas \inhouse \ charges €125.00, marking a 40\% increase. Work outside the agreement is billed at €64.90 per hour by \private, but at €135.00 by \inhouse, indicating a 108\% difference. Assistant work costs €60.00 per hour at \private \ and €89.00 at \inhouse, reflecting a 48\% difference. These differences highlight the higher costs associated with \inhouse's services, raising questions about efficiency.

\section{Discussion}

In short, business acquisition over \inhouse's financial and payroll services improved the following practices and reduced pricing:
\begin{itemize}
    \item Transition to Monthly Subscription Model: Used hours per client decreased by 17.3\% to 56.3\%.
    \item Enhanced Work Practices: Overall billing increased by 22\%, and employee efficiency improved by 82.5\%. 
    \item Automation and Skill Updating: Client service efficiency increased by 55\%.
    \item Furthermore, hourly rates showed significant differences following the acquisition. \private's hourly rates and services are on average 41\% less expensive.
    \item Differences in the Collective Bargaining Agreement: ERTO agreement is over 15\% more cost-effective. \private \ employees work 9\% more hours, resulting in a 15\% lower hourly labour cost.

\end{itemize}

\inhouse, as a publicly funded in-house entity, is expected to deliver services efficiently and cost-effectively to its public body owners. However, the financial analysis raises concerns about whether this objective is being met. The comparison between the top-down and bottom-up analysis shows clear differences between \inhouse\ and \private. \inhouse\ has significantly higher labor costs relative to its revenue, with \private\ being 41\% more cost-effective in this area. Combined with weaker liquidity, \inhouse\ faces higher financial risk.

 \inhouse's higher labour costs, exemplified by its hourly rates for various services, suggest inefficiencies that contradict the goal of reducing operational expenses. For instance, payroll processing costs at \inhouse \ are significantly higher than those at \private, and similar patterns are observed in other service categories such as bookkeeping, expert work, and assistant services.

These elevated costs directly impact \inhouse's profitability and cost-effectiveness. Despite \inhouse's high revenue and substantial return on equity (ROE), its contribution margin and net profit margin are considerably lower than those of \private. This indicates that \inhouse \ is less efficient at converting revenue into profit, which is a critical measure of operational efficiency. Additionally, \inhouse's quick ratio, while adequate, is significantly lower than \private's, suggesting less liquidity and potentially greater financial vulnerability.

Given that \inhouse \ operates on public funds, these inefficiencies are particularly concerning. The higher costs could mean that public bodies are not getting the best value for their money, which undermines the rationale for using an in-house entity. Public agencies aim to optimize their budgets to maximize the benefits for the community, and paying more for services that could be obtained more cost-effectively from the market contradicts this goal.

The decision for public agencies, then, hinges on a comparative assessment of costs and benefits. If market solutions can provide the same services at lower costs and with higher efficiency, it would be financially prudent for public bodies to consider outsourcing rather than relying solely on their in-house entity. The evidence from the financial analysis suggests that \private, with its lower costs and higher efficiency, could potentially offer more cost-effective solutions compared to \inhouse.

In conclusion, while \inhouse's role as a publicly funded in-house entity is to provide services efficiently and cost-effectively, the financial data indicates that it may not be fulfilling this mandate effectively. Public agencies should therefore critically evaluate whether continuing with an in-house entity is the best option or if procuring solutions from the market would better serve their financial and operational goals. This evaluation should be based on a thorough comparison of costs, efficiencies, and the overall value provided by potential service providers.


\textbf{Limitations.} While the bottom-up analysis provides useful comparisons of labor costs and highlights important factors such as automation that impact efficiency, it remains challenging to fully understand the operational effectiveness behind the numbers. This limitation suggests that, while current data offer valuable insights, there is still room for improvement in capturing and interpreting the true efficiency of operations. Future research could aim to better articulate these underlying factors to provide a more comprehensive view of organizational performance.

\section{Conclusion}

Using the case study method with both top-down and bottom-up approaches has provided a detailed understanding of the financial and operational impacts of ownership changes from in-house to private companies.

From the top-down analysis, it became clear that \private \ is financially more stable, with stronger profitability and liquidity. \private \ manages costs more efficiently, with a more robust financial structure. On the other hand, \inhouse, though effective in sales and demonstrating a high return on equity, has a weaker financial structure and lower profitability compared to \private.

The bottom-up analysis revealed that after the acquisition, efficiency improved by 44\%. A significant finding was that \inhouse's hourly rates are, on average, 57\% higher than \private’s when excluding payroll services, and 90\% higher when including them. Additionally, \inhouse’s labour costs are notably higher due to its collective bargaining agreement, resulting in 8\% higher annual salary costs and a 15\% higher hourly labour cost. This is driven by higher salaries, holiday compensation, and replacement costs, combined with shorter working hours.

This dual approach allowed us to see how privatization can lead to better cost management and operational efficiency in the private sector, while publicly funded entities like \inhouse \ face challenges aligning their costs with public sector efficiency goals. The findings suggest that effective financial oversight and renegotiation of labour agreements may be necessary to improve cost-effectiveness in publicly owned firms.

%
%


\begin{thebibliography}{6}
%

\bibitem {tutki:hallintoa}
Explore administration. 2024. \url{https://www.tutkihallintoa.fi/julkiset-hankinnat/hankintojen-arvo/#:~:text=Julkisten%20hankintojen%20arvo%20on%20vuodessa,hankkii%20tavaroita%20ja%20palveluja%20vuosittain.}
\bibitem {hankinta:laki}
The Public Procurement Act. 1397/2016. Act on Public Procurement and Concession Contracts. Translation from Finnish. Legally binding only in Finnish and Swedish. Ministry of Economic Affairs and Employment, Finland

\bibitem {inhouse:maara}
Miettinen, V. (2024, January 19). Kuntaliitto vetoaa hallitukseen: Inhouse-yhtiöiden sääntely arvioitava uudelleen – “Kustannukset voivat ylittää hyödyt” | Kuntalehti. Kuntalehti. Retrieved September 8, 2024. https://kuntalehti.fi/uutiset/talous/kuntaliitto-vetoaa-hallitukseen-inhouse-yhtioiden-saantely-arvioitava-uudelleen-kustannukset-voivat-ylittaa-hyody/.

\bibitem {pyykkonen:saastokeinot}
Pyykkönen, J., Halonen, K. M., Tukiainen, J., \& Parviainen, A. (2023). Selvitys julkisen hankintojen säästökeinoista. https://vm.fi/documents/10623/150718668/Selvitys+julkisen+hankintojen+s

\bibitem {KKV:marketcourt}
Market Court: Sarastia not an in-house entity of Vantaa and Kerava Wellbeing Services County. Released 15.3.2024.
https://www.kkv.fi/en/current/press-releases/market-court-sarastia-not-an-in-house-entity-of-vantaa-and-kerava-wellbeing-services-county/

\bibitem {halonen:inhouse}
Halonen, K.-M. (2024, June 13). In-house-sääntelystä ja syistä sidosyksikkökehityksen taustalla (KKV:n Katsauksia 4/2024). Kilpailu- Ja Kuluttajavirasto. Retrieved September 8, 2024, from https://www.kkv.fi/tutkimus-ja-vaikuttaminen/julkaisut/katsaukset/in-house-saantelysta-ja-syista-sidosyksikkokehityksen-taustalla-kkvn-katsauksia-4-2024/

\bibitem {government:programme}
A strong and committed Finland. Programme of Prime Minister Petteri Orpo's Government. Publications of the Finnish Government 2023:60. 
\url{https://valtioneuvosto.fi/en/governments/government-programme#/}

\bibitem {miemec:2019}
Miemiec, W. (2019). The application of in-house procurement by municipalities in municipal services management.

\bibitem {hausmann2013house}
Hausmann, F. L., \& Queisner, G. (2013). In-House Contracts and Inter-Municipal Cooperation–Exceptions from the European Union Procurement Law Should be Applied with Caution. European Procurement \& Public Private Partnership Law Review, 8(3), 231-237.

\bibitem {baciu2015horizontal}
Baciu, I., \& Dragoş, D. C. (2015). Horizontal in-house transactions vs. vertical in-house transactions and public-public cooperation. European Procurement \& Public Private Partnership Law Review, 10(4), 254-272.

\bibitem {myers:2008}
Myers, M. D. (2008). Qualitative Research in Business and Management. http://www.gbv.de/dms/zbw/574672206.pdf

\bibitem {hankinta:direktiivi}
Directive 2014/24/EU of the European Parliament and of the council of 26 February 2014 on public procurement and repealing Directive 2004/18/EC. Official Journal of the European Union.

\bibitem {MAO:vantaakerava}
MAO:154/2024. Finnish Market Court. Case Law. 

\bibitem {Cooper:1992}
Cooper, R., \& Kaplan, R. S. (1992). Activity-based systems: Measuring the costs of resource usage. Accounting horizons, 6(3).

\bibitem {Laine:2012}
Laine, T., Paranko, J., \& Suomala, P. (2012). Management accounting roles in supporting servitisation: implications for decision making at multiple levels. Managing Service Quality: An International Journal, 22(3), 212-232.  

\bibitem {onvire:tilinp}
Onvire tilinpäätös 1.1.2022–31.12.2022. https://numera.fi/wp-content/uploads/2023/10/onviretilinpaatos2022-1.pdf
\url{doi:10.1007/11823285_121}

\bibitem {MA:lindholm}
Lindholm, A., Laine, T. J., \& Suomala, P. (2017). The potential of management accounting and control in global operations. Journal of Service Theory and Practice. 27(2), 496-514.


\bibitem {jaaskelainen:kkv}
Jääskeläinen, J., Saastamonen, A., Väättänen, A., \& Hiilamo, T. (2024, September 4). Tutkimus 10 prosentin omistusosuusvaatimuksen vaikutuksista sidosyksikkömarkkinaan (Tutkimusraportteja 2/2024, päivitetty versio 9/2024). Kilpailu- Ja Kuluttajavirasto. Retrieved September 8, 2024, from https://www.kkv.fi/tutkimus-ja-vaikuttaminen/julkaisut/tutkimusraportit/tutkimus-10-prosentin-omistusosuusvaatimuksen-vaikutuksista-sidosyksikkomarkkinaan-tutkimusraportteja-2-2024/

\bibitem {shulver:2005}
Shulver, M. (2005). Operational loss and new service design. International Journal of Service Industry Management, 16(5), 455-479.

\bibitem {kastalli:2013}
Kastalli, I. V., Van Looy, B., \& Neely, A. (2013). Steering manufacturing firms towards service business model innovation. California management review, 56(1), 100-123.

\bibitem {KKV:esitys}
KKV vie Vantaan ja Keravan hyvinvointialueen Sarastia-hankinnan markkinaoikeuden arvioitavaksi. (2023, May 23). Kilpailu- Ja Kuluttajavirasto. https://www.kkv.fi/ajankohtaista/tiedotteet/kkv-vie-vantaan-ja-keravan-hyvinvointialueen-sarastia-hankinnan-markkinaoikeuden-arvioitavaksi/


\bibitem {EK:2024}
Elinkeinoelämän järjestöt: KKV:n ratkaisu hankintalain kiertämisestä tarpeellinen Suomen kasvulle: ”Usko markkinatalouteen vahvistui” - Elinkeinoelämän keskusliitto. (2023, May 23). Elinkeinoelämän Keskusliitto. https://ek.fi/ajankohtaista/tiedotteet/elinkeinoelaman-jarjestot-kkvn-ratkaisu-hankintalain-kiertamisesta-tarpeellinen-suomen-kasvulle-usko-markkinatalouteen-vahvistui/

\end{thebibliography}
\end{document}